\documentclass[]{article}
\usepackage[T1]{fontenc}                
\usepackage[utf8]{inputenc}             
\usepackage{amssymb,amsthm,amsmath}    
\usepackage{enumerate}
\usepackage{paralist}
\usepackage{authblk}
\usepackage{graphicx}
\usepackage{enumerate}
\usepackage{array}
\usepackage{tikz}
\usetikzlibrary{calc}
\usetikzlibrary{decorations.pathreplacing}

\newcommand{\N}{\mathbb{N}}
\newtheorem{theorem}{Theorem}
\newtheorem{lemma}{Lemma}

\newtheorem*{prob*}{Problem}
\newtheorem{prop}{Proposition}
\newtheorem*{prop*}{Proposition}

\newtheorem*{observ*}{Observation}
\title{Diagonalization and Undecidability in Semi-Thue Systems}
\author{Vesa Halava, Reino Niskanen}

\begin{document}

\maketitle

\begin{abstract}
We revisit classical diagonalization argument in the setting of semi-Thue systems, viewing computation as nondeterministic rewriting rather than sequential machine execution. Semi-Thue systems naturally generate derivation trees, making existential acceptance, reachability, and infinite derivations fundamental computational notions.

We formulate acceptance and decision directly in terms of derivation trees and study the associated assertion, word, and infiniteness problems. We prove that the complements of the assertion and word problems are not recursively enumerable by direct Post-style diagonal arguments. For the infiniteness problem, we give a constructive diagonalization based on a fixed universal semi-Thue system. Unlike the classical approach, the diagonal construction is carried out entirely within the rewriting formalism: the universal system remains an ordinary nondeterministic semi-Thue system while simulating encoded semi-Thue computations, including self-reference.

These results show that undecidability in semi-Thue systems arises naturally from universality and self-reference in nondeterministic rewriting, rather than from machine states or control mechanisms. This perspective clarifies the relationship between rewriting systems, recursive enumerability, and classical diagonal arguments originating in the work of Post and Turing.

\end{abstract}

\section{Introduction}

The unifying method behind the earliest undecidability results is diagonalization over effective descriptions of computational processes, conceptually rooted in Cantor’s diagonal argument for the uncountability of the real numbers. This idea appears already in Turing’s analysis of computable numbers, where the notion of circular-freeness is used to separate computable from non-computable real numbers by an explicit diagonal construction~\cite{Turing1936}. It is subsequently recast as an explicit decision problem in the formulation of the halting problem for Turing machines, as systematically developed by Davis~\cite{Davis1958}. Post’s separation of recursive and recursively enumerable sets likewise relies on diagonal reasoning, carried out in the formalism of normal systems, an early nondeterministic rewriting model~\cite{Post1944}.  

These results are standardly presented within machine-based formalisms and serve as canonical instances of undecidability obtained by diagonalization.

In this paper we revisit diagonalization in a markedly different computational model, namely
\emph{semi-Thue systems}, whose computations are inherently nondeterministic and tree-like,
and whose derivations resemble proof search in a formal logical system rather than sequential
machine execution. This perspective is already implicit in Post’s use of normal systems
as a rewriting formalism for recursively enumerable sets~\cite{Post1944}, where diagonalization
is carried out over derivation trees rather than over machine–input pairs. Our contribution is
to make this perspective explicit and systematic, and to formulate diagonal arguments directly
at the level of semi-Thue derivations and their acceptance modes.

The first motivation of this work is to recall that diagonalization is not tied
to the specific operational structure of Turing machines, but can be carried out
directly in rewriting-based models of computation.
Semi-Thue systems are among the simplest and most natural such models.
They lack states, instruction counters, or explicit control mechanisms,
yet they are computationally universal.
This raises a conceptual question: how do classical undecidability arguments
manifest themselves in a formalism where computation is given by a branching
derivation tree rather than a linear run?
Indeed, recall that universal Turing machines are deterministic with several papers investigating the boundaries of the universality with respect to the number of states and size of the input alphabet \cite{Shannon56, Rogozhin96,WN09}.
This makes it seem that the computational universality arises from computational devices that have sufficiently rich underlying structure.

From a contemporary perspective, tree-like and nondeterministic modes of
computation are increasingly common.
On one hand, they are present in SAT and SMT solvers, where an assignment can be nondeterministically restarted by the solver if the current computation takes too long.
On the other hand, many modern computational systems, including large language models and related
neural-symbolic architectures, operate operationally as forms of guided,
branching search rather than as single linear executions.
While such systems are not intended as formal models of effective computation,
their prevalence highlights the naturalness of tree-structured computation
as a basic computational paradigm.

Our primary technical goal is to give direct proofs of undecidability for natural problems of semi-Thue systems, treating these problems themselves as fundamental sources of undecidability, rather than establishing their undecidability by reduction from previously known undecidable problems for Turing machines.
In particular, we study the \emph{assertion problem}:
given a semi-Thue system and a designated start word, does there exist a finite
derivation in which a distinguished symbol occurs?
This problem captures the recursively enumerable nature of semi-Thue computation
in a particularly direct way.
While assertion is semi-decidable by systematic enumeration of derivations,
we show that it is not decidable; moreover, its complement is not recursively
enumerable.

Closely related is the classical \emph{word problem} for semi-Thue systems:
given a system and two words, does there exist a finite derivation transforming
one word into the other?
Like assertion, the word problem is naturally formulated in existential terms
over derivation trees and is therefore recursively enumerable.
It provides a canonical notion of reachability in rewriting-based models of
computation and serves as a standard benchmark problem for undecidability
results in this setting.

We also consider non-termination properties, such as the existence of an infinite
derivation.
These correspond to non-halting or circular-freeness phenomena in Turing machine
models and lie outside the class of recursively enumerable properties.
We show that the undecidability of non-termination can likewise be obtained by
a direct diagonalization argument using a universal semi-Thue system.

Historically, the approach taken here is closely related to that of Emil Post
\cite{Post1943,Post1944}.
In his work on normal systems, Post emphasized generative formalisms and treated
computation as the production of assertions rather than as machine execution.
Acceptance was represented by the appearance of a distinguished symbol.
Our treatment follows this tradition.

At this point it is important to address the notion of control.
What is commonly called control in a Turing machine—its states and halting states—
is not intrinsic supervision of computation but a convention for communicating
the result to an external observer.
The machine itself merely follows local rules.
The essential difference between Turing machines and semi-Thue systems is that
the former compress an unbounded computation into a single finite observable
configuration, whereas the latter distribute the result over a potentially infinite
set of derivable words.

\section{Semi-Thue Systems and Derivations}

\subsection{Semi-Thue systems}

Let $A$ be a finite alphabet. A \emph{word} over $A$ is a finite
sequence of symbols from $A$. We denote by $A^*$ the set of all words
over $A$; equivalently, $A^*$ is the free monoid generated by $A$ under
concatenation. The empty word is denoted by $\varepsilon$.

If
\(  w=a_1a_2\cdots a_n \)
with $a_i\in A$, then the \emph{length} of $w$ is $n$, and it is denoted
by $|w|$. In particular, $|\varepsilon|=0$.

A word $u\in A^*$ is called a \emph{factor} of a word $w\in A^*$ if
there exist words $p,q\in A^*$ such that
$w=puq.$  
In this case we also say that $u$ occurs in $w$. If $p=\varepsilon$,
then $u$ is a \emph{prefix} of $w$, and if $q=\varepsilon$, then $u$ is a
\emph{suffix} of $w$.

A \emph{semi-Thue system} is a pair
$S=(A,R)$, 
where $A$ is a finite alphabet and $R\subseteq A^*\times A^*$ is a
finite set of rewriting rules. A rule $(u,v)\in R$ will usually be
written as $u\to v.$

For words $x,y\in A^*$, we write $x\to_S y$
if there exist words $p,q\in A^*$ and a rule $u\to v$ in $R$ such that
\[
        x=puq
        \quad\text{and}\quad
        y=pvq.
\]
Thus a single rewriting step consists of replacing one occurrence of
the left-hand side of a rule by its right-hand side, leaving the
surrounding context unchanged.

The reflexive and transitive closure of $\to_S$ is denoted by
$\to_S^*$. Hence $ x\to_S^* y$ means that $y$ is obtainable from $x$ by a finite sequence of rewriting
steps, possibly of length zero. Such a sequence of rewriting steps is called a \emph{finite derivation}. In other words, a finite derivation in  $S$ is a finite sequence of words
\[
        w_0,w_1,\ldots,w_k
\]
such that $w_i\to_S w_{i+1}$ for all $i<k$.
We also write this as
\[
        w_0\to_S w_1\to_S\cdots\to_S w_k.
\]

An \emph{infinite derivation} in $S$ is an infinite sequence of words
\[
        w_0,w_1,w_2,\ldots
\]
such that $w_i\to_S w_{i+1}$  for all $ i\in\N$.

A word $w\in A^*$ is called \emph{irreducible} with respect to $S$ if
there is no word $w'$ such that$ w\to_S w'$.
Equivalently, no left-hand side of a rule of $S$ occurs as a factor of
$w$.

A derivation
\[
        w_0\to_S w_1\to_S w_2\to_S\cdots
\]
is called \emph{maximal} if either it is infinite, or it is finite
\[
        w_0\to_S w_1\to_S\cdots\to_S w_k
\]
and the last word $w_k$ is irreducible.

Thus a finite maximal derivation is a derivation which cannot be extended. In other words, maximal derivation either continues forever, or it stops because no rule is applicable.

A brief remark on the encodings used throughout the paper.
We fix an effective coding of finite alphabets,
finite semi-Thue systems, words, and finite tuples of such objects by
words over a fixed finite alphabet. We write these codes as
\[
        \langle S\rangle,\qquad
        \langle S,w\rangle,\qquad
        \langle S,u,v\rangle,
\]
and so on.

The particular coding is irrelevant, provided that it is effective in
the following sense. From a code $\langle S\rangle$ one can effectively
recover the finite alphabet and the finite list of rules of $S$; from
codes of finite objects one can effectively form codes of finite tuples;
and every effective transformation of semi-Thue systems and words
induces a computable transformation of their codes.

When the fixed universal semi-Thue system $\mathcal U$ is used, we shall
use a more specific encoding \(\varphi\) that is described in Section~\ref{sec:universal-sT}.

Thus there are two levels of coding in the paper. The notation
$\langle S,w\rangle$ denotes an arbitrary effective external code of an
instance of a decision problem, whereas $\varphi(S,w)$ denotes the
specific internal start word used by the fixed universal semi-Thue
system $\mathcal U$.

\subsection{The word and termination problems}

The \emph{word problem} for semi-Thue systems is the following decision
problem: given a semi-Thue system $S=(A,R)$ and two words $u,v\in A^*$,
decide whether
\[
        u \to_S^* v.
\]
In other words, the word problem asks whether $v$ is reachable from $u$
by finitely many applications of the rules of $S$.

The same problem can also be formulated as a membership problem for a
set of encoded instances. Define
\[
        \mathrm{WP}
        =
        \{\langle S,u,v\rangle \mid u\to_S^* v\}.
\]
Here $\langle S,u,v\rangle$ denotes an effective encoding of the finite
object consisting of the system $S$ and the two words $u,v$.

Thus, given a triple $\langle S,u,v\rangle$, the task in the word problem
is to decide whether
\[
        \langle S,u,v\rangle\in \mathrm{WP}.
\]

Let $S$ be a semi-Thue system and let $w$ be a word. We say that $S$
\emph{terminates from} $w$ if every derivation beginning with $w$ is
finite. Equivalently, there is no infinite derivation
\[
        w=w_0\to_S w_1\to_S w_2\to_S\cdots.
\]
It is now easy to define the \emph{termination problem} for the semi-Thue systems: given a semi-Thue system $S=(A,R)$ and a word $u$, decide whether $S$ terminates on $u$. 
One can easily see how the termination problem for semi-Thue systems is analogous to the halting problem of the Turing machines; see \cite{Davis1958}.

We define the set
\[
        \mathrm{INF}
        =
        \{\langle S,w\rangle
        \mid
        \text{there exists an infinite derivation from } w
        \text{ in } S\}.
\]
Thus $\langle S,w\rangle\in\mathrm{INF}$ means that $S$ does not
terminate from $w$ in the existential sense: at least one infinite
branch exists in the derivation tree rooted at $w$.

The complementary universal termination property is now defined by 
\[
        \mathrm{TERM}
        =
        \{\langle S,w\rangle
        \mid
        \text{every derivation from } w \text{ in } S
        \text{ is finite}\}.
\]
Thus $\mathrm{TERM}=\overline{\mathrm{INF}}$.

\subsection{A fixed universal semi-Thue system}
\label{sec:universal-sT}

For the diagonal argument concerning infinite derivations we use a fixed
universal semi-Thue system. The system is the one constructed by
Halava, Matiyasevich and Niskanen~\cite{HalavaMatiyasevichNiskanen2017}. It is universal with
respect to the existence of infinite derivations. We recall here only
the part of the construction that is needed below: the coding of an
arbitrary semi-Thue system and its initial word into a single initial
word of the fixed universal system. The construction is related to Tseitin's famous seven-rule semi-Thue system with undecidable word problem, see~\cite{Nyberg-Brodda2026} for a survey on the topic.

The universal construction is stated for semi-Thue systems whose rule
words are nonempty. This entails no loss of generality for the
termination/finiteness problem as can be seen by the following Lemma.

\begin{lemma}\label{lem:remove-empty-sides}
There is an effective construction which, given a semi-Thue system
$S=(A,R)$ and a word $w\in A^*$, produces a semi-Thue system
$S^\sharp=(A^\sharp,R^\sharp)$, where \(A^\sharp=A\cup\{\sharp\}\), and a word
$w^\sharp\in (A^\sharp)^*$ such that no rule of $S^\sharp$ has the
empty word as either of its sides, and
\[
        \text{there exists an infinite derivation from } w
        \text{ in } S
\]
if and only if
\[
        \text{there exists an infinite derivation from } w^\sharp
        \text{ in } S^\sharp.
\]
\end{lemma}
The simplest way of proving the above lemma is to add a new symbol $\sharp\notin A$ before, after, and between the symbols of the words, and to replace each occurrence of $\varepsilon$ in the rules of $R$ by $\sharp$.

Thus, when applying the fixed universal system to the infiniteness
problem, we may assume without loss of generality that all rule words are
nonempty. In the following description, let
\(
        S=(A,R)
\)
be a semi-Thue system, where
\(
        A=\{a_1,\ldots,a_m\}
\)
and
\(
        R=\{(u_1,v_1),\ldots,(u_n,v_n)\},
\)
with $u_i\neq\varepsilon$ and         $ v_i\neq\varepsilon$
for all $i=1,\ldots,n$.

We first encode words over $A$ by words over the two-letter alphabet
$\{a,b\}$. If
\[
        w=a_{i_1}a_{i_2}\cdots a_{i_k}\in A^*,
\]
we define
\[
        \sigma(w)=ab^{i_1}ab^{i_2}\cdots ab^{i_k}.
\]
We also put
\[
        \widetilde{\sigma}(w)=\sigma(w)a.
\]
Thus each letter $a_i$ is encoded by the block $ab^i$, and the final
letter $a$ marks the end of the encoded word.

The fixed universal semi-Thue system defined in \cite{HalavaMatiyasevichNiskanen2017} is denoted by
\(
        \mathcal U=(\Sigma,R_{\mathcal U}),
\)
where
\(
        \Sigma=\{a,b,c,d,e_1,e_2,e_3,e_4\}.
\)
We do not present the full rule set of \(\mathcal{U}\).
Instead we highlight the key purposes of different letters.

The letters $a,b$ are used for the encoded current word, whereas the
letters $c,d$ are used for the encoded rule system. We define morphism
\[
        \alpha:\{a,b\}^*\to\{c,d\}^*
\]
by
\[
        \alpha(a)=c,\qquad \alpha(b)=d,
\]
In other words, $\alpha$ changes the alphabet $\{a,b\}$ into the alphabet
$\{c,d\}$.
Together with \(\sigma\) it encodes rules \(R\subseteq A^*\times A^*\) into the alphabet \(\{c,d\}\).
To be more precise, for a word
\[
        w=a_{i_1}a_{i_2}\cdots a_{i_k}\in A^*
\]
we define
\[
        \tau(w)=\alpha(\sigma(w))
        =
        cd^{i_1}cd^{i_2}\cdots cd^{i_k}.
\]
The rules of $S$ are encoded by setting
\[
        \tau((u_i,v_i))=cc\,\tau(u_i)c\tau(v_i)
\]
and
\[
        \tau(R)
        =
        \tau((u_1,v_1))\cdots \tau((u_n,v_n))cc.
\]
Since the rule words are assumed to be nonempty, the word $\tau(R)$
determines the finite list of rules uniquely.

Now the pair consisting of the semi-Thue system $S=(A,R)$ and the
initial word $w$ is encoded into the single word
\[
        \varphi(S,w)=\tau(R)e_1\widetilde{\sigma}(w)
\]
over the alphabet $\Sigma$ of the fixed system $\mathcal U$. This is the specific internal encoding used by the fixed universal
semi-Thue system $\mathcal U$. It should be distinguished from the
external code $\langle S,w\rangle$ of an instance of a decision problem.

The role of the marker $e_1$ is to separate the encoded rule table
$\tau(R)$ from the encoded current word $\widetilde{\sigma}(w)$.
During a simulation step, the rules of $\mathcal U$ move the markers
$e_1,e_2,e_3,e_4$ through the word in order to compare an encoded
left-hand side of a rule with an occurrence in the current word, replace
it by the corresponding right-hand side, and finally restore the word to
the form
\[
        \tau(R)e_1\widetilde{\sigma}(w').
\]
Thus one simulated rewriting step of $S$ corresponds to a finite
derivation in $\mathcal U$. The following proposition follows from the proof of the  next theorem from \cite{HalavaMatiyasevichNiskanen2017}.

\begin{prop}\label{prop:universal-finite-simulation}
Let $S=(A,R)$ be a semi-Thue system whose rule words are nonempty, and
let $\mathcal U$ and $\varphi$ be as above. If
\[
        w_0\to_S w_1\to_S\cdots\to_S w_n
\]
is a finite derivation in $S$, then
\[
        \varphi(S,w_0)
        \to_{\mathcal U}^*
        \varphi(S,w_1)
        \to_{\mathcal U}^*
        \cdots
        \to_{\mathcal U}^*
        \varphi(S,w_n).
\]
\end{prop}

Consequently, every infinite derivation
\[
        w=w_0\to_S w_1\to_S w_2\to_S\cdots
\]
in $S$ gives rise to an infinite derivation
\[
        \varphi(S,w_0)
        \to_{\mathcal U}^*
        \varphi(S,w_1)
        \to_{\mathcal U}^*
        \varphi(S,w_2)
        \to_{\mathcal U}^*
        \cdots
\]
in $\mathcal U$.

The essential property of $\mathcal U$ is the converse direction as
well. Although a derivation in $\mathcal U$ may pass through many
intermediate words which are not themselves of the form
$\varphi(S,z)$, every infinite derivation from $\varphi(S,w)$ yields an
infinite derivation from $w$ in the simulated system $S$.

The theorem was proved in \cite{HalavaMatiyasevichNiskanen2017}.

\begin{theorem}\label{thm:universal-st-inf}
For every semi-Thue system $S=(A,R)$ and every word
$w\in A^*$, one can effectively construct a word
\(
        \varphi(S,w)\in\Sigma^*
\)
with the following property:
\[
        \text{there exists an infinite derivation from } w
        \text{ in } S
\]
if and only if
\[
        \text{there exists an infinite derivation from } \varphi(S,w)
        \text{ in } \mathcal U.
\]
\end{theorem}

Equivalently,
\[
        \langle S,w\rangle\in\mathrm{INF}
        \quad\Longleftrightarrow\quad
        \langle \mathcal U,\varphi(S,w)\rangle\in\mathrm{INF}.
\]
Thus the existence of an infinite derivation for an arbitrary
semi-Thue system can be represented as the existence of an infinite
derivation in one fixed semi-Thue system.

This fixed-system formulation is the semi-Thue analogue of using a
universal Turing machine. The input semi-Thue system is no longer part
of the rule set being executed; it is part of the initial word. This is
the form of universality needed for the diagonal constructions below.

\section{Acceptance, Assertion, and Decision}

\subsection{Existential acceptance}

Let $S=(A,R)$ be a semi-Thue system, and let $Y\in A$ be a distinguished
symbol. The symbol $Y$ is called the \emph{acceptance symbol}.

We say that $S$ \emph{accepts} a word $w\in A^*$ if there exists a word
$u\in A^*$ such that
\(
        w\to_S^* u
\)
and $Y$ occurs in $u$.
Equivalently,
\[
        S \text{ accepts } w
        \quad\Longleftrightarrow\quad
        \exists u\in A^*
        \bigl(w\to_S^*u
        \text{ and } Y \text{ occurs in } u\bigr).
\]

This is an existential notion of acceptance. It does not require that
all derivations from $w$ produce $Y$, nor does it require that a word
containing $Y$ be irreducible. Acceptance means only that some finite
branch of the derivation tree reaches a word in which the distinguished
symbol $Y$ occurs.

This use of a distinguished symbol is close in spirit to Post's use of
normal systems. In Post's formulation, computation is presented
generatively: a system produces assertions rather than executing a
linear sequence of machine instructions. In particular, recursively
enumerable sets of positive integers may be represented by normal
systems over an alphabet such as $\{1,b\}$, where the integer $n$ is
represented by the word $1^n$, and the occurrence of the distinguished
symbol $b$ serves to mark a generated assertion. Thus acceptance is not
an internal act of decision by the system itself, but the external
recognition that a word of the required form has appeared in the
derivation tree. Our acceptance symbol $Y$ plays the same formal role:
it is a marker whose occurrence witnesses membership by the existence of
a finite derivation. This is close to Post's formulation of recursively enumerable sets by
normal systems, where generated words are treated as assertions; see
Urquhart~\cite{Urquhart2009Post}.

The \emph{assertion problem} for semi-Thue systems is the set
\[
        \mathrm{ASSERT}
        =
        \{\langle S,w\rangle
        \mid
        S \text{ accepts } w\}.
\]
Thus $\langle S,w\rangle\in\mathrm{ASSERT}$ if and only if there exists
a finite derivation
\(        w\to_S^* u
\)
such that $Y$ occurs in $u$.

The assertion problem is the natural reachability problem associated
with existential acceptance. It asks whether the derivation tree rooted
at $w$ contains at least one accepting node.

\subsection{Decision by semi-Thue systems}
For the purpose of discussing decision rather than mere recognition, it
is useful to distinguish two special symbols
$Y$ and $N$,
where $Y$ represents a positive verdict and $N$ represents a negative
verdict.

Let $S=(A,R)$ be a semi-Thue system with $Y,N\in A$. A word $u\in A^*$
is called \emph{verdict-bearing} if either $Y$ occurs in $u$ or $N$
occurs in $u$. We write
\[
        \mathrm{Verd}
        =
        \{u\in A^*
        \mid
        Y \text{ occurs in } u
        \text{ or }
        N \text{ occurs in } u\}.
\]

The symbols $Y$ and $N$ have no intrinsic computational meaning. They
become acceptance and rejection symbols only through the external
interpretation assigned to them.

Let $L\subseteq B^*$ be a language over a finite alphabet $B$. To say
that a semi-Thue system $S=(A,R)$ recognizes or decides $L$, we need an
effective way of associating to each input word $x\in B^*$ a start word
over the alphabet $A$.

We therefore fix a computable map
\[
        \mathrm{start}\colon B^*\to A^*.
\]
The word $\mathrm{start}(x)$ is the initial word from which the
derivation of $S$ on input $x\in B^*$ begins.

In many constructions the map $\mathrm{start}$ is simply an encoding of
$x$ into the alphabet of $S$. We keep it explicit in order to separate
the input representation from the rewriting rules themselves.

Assume further that \(S\) has 
distinguished symbols $Y,N\in A$.
We say that $S$ is \emph{existentially correct} for $L$ if, for every
input word $x\in B^*$, the following two conditions hold:

\begin{enumerate}
    \item if $x\in L$, then there exists a derivation
    \[
            \mathrm{start}(x)\to_S^* u
    \]
    such that $Y$ occurs in $u$, and there is no derivation
    \[
            \mathrm{start}(x)\to_S^* v
    \]
    such that $N$ occurs in $v$;

    \item if $x\notin L$, then there exists a derivation
    \[
            \mathrm{start}(x)\to_S^* u
    \]
    such that $N$ occurs in $u$, and there is no derivation
    \[
            \mathrm{start}(x)\to_S^* v
    \]
    such that $Y$ occurs in $v$.
\end{enumerate}

Thus existential correctness says that the correct verdict is reachable
and the wrong verdict is not reachable. It does not yet say that every
maximal derivation reaches a verdict. There may still be derivations
which avoid both $Y$ and $N$ forever, or finite maximal derivations which
stop without producing either symbol.

We say that $S$ is \emph{verdict-total} with respect to $\mathrm{start}$
if, for every input $x\in B^*$, every maximal derivation from
$\mathrm{start}(x)$ reaches a verdict. That is, for every maximal
derivation
\[
        w_0,w_1,w_2,\ldots
\]
with
\[
        w_0=\mathrm{start}(x),
\]
there exists an index $i$ such that
\[
        w_i\in \mathrm{Verd}.
\]

Equivalently, there is no maximal derivation from $\mathrm{start}(x)$
which avoids both $Y$ and $N$ forever.

Verdict totality is the rewriting-theoretic analogue of the requirement
that all computation branches eventually produce an observable outcome.

Let $L\subseteq B^*$ be a language. A semi-Thue system $S=(A,R)$ with
distinguished verdict symbols $Y,N\in A$ \emph{decides} $L$ if there
exists a computable input map
\[
        \mathrm{start}\colon B^*\to A^*
\]
such that:

\begin{enumerate}
    \item $S$ is existentially correct for $L$ with respect to
    $\mathrm{start}$;

    \item $S$ is verdict-total with respect to $\mathrm{start}$.
\end{enumerate}

Thus decision consists of two separate requirements. First, the system
must be correct: the right verdict must be reachable and the wrong
verdict must be unreachable. Second, the system must be total from the
observer's point of view: every maximal derivation must eventually reach
some verdict.

Existential acceptance alone is therefore a notion of recognition, not
decision. Decision requires that nondeterministic branching cannot hide
the answer in a verdict-avoiding branch.

The distinction between recognition and decision is central in what
follows. A semi-Thue system recognizes a language by producing a positive
witness on at least one branch of a derivation tree. This is the
semi-Thue counterpart of recursive enumerability.

Decision is stronger. It requires both positive and negative verdicts,
and it requires that every maximal derivation reveal one of them. In
this sense, verdict totality is a semantic determinization condition:
the system may remain nondeterministic internally, but the nondeterminism
must not prevent the observer from obtaining a final answer.

The undecidability results below are first proved at the level of
existential acceptance and infinite derivations. The later discussion of
decision explains how these non-recursive-enumerability results imply
classical undecidability.

\section{Assertion and Recursive Enumerability}

\subsection{RE sets via semi-Thue acceptance}

A language $L\subseteq A^*$ is \emph{recursively enumerable} if there exists a semi-Thue
system $S_L$ such that for all $w$,
\[
w\in L \quad\Longleftrightarrow\quad S_L \text{ accepts } w.
\]
Post's original formulation of recursively enumerable sets by normal
systems is given in~\cite{Post1944}; see also the reprints in
Davis~\cite[pp.~305--337]{Davis1965Undecidable} and in Post's collected
works~\cite[pp.~461--494]{Post1994Collected}. In Post's formulation,
computation is presented generatively: a system produces assertions
rather than executing a linear sequence of machine instructions. Thus
membership is witnessed by the appearance of a generated word of the
required form.

This viewpoint is close to the use of the acceptance symbol $Y$ above.
The occurrence of $Y$ is not an internal act of decision by the
semi-Thue system itself, but an externally observed marker witnessing
that a successful finite derivation has been produced. The diagonal
reading of Post's construction is discussed by
Urquhart~\cite{Urquhart2009Post}, who describes Post's enumeration of
normal systems and the associated diagonal set.

It is straightforward to see that the assertion problem is recursively enumerable. Indeed, given a
semi-Thue system $S=(A,R)$ and a word $w\in A^*$, one can enumerate the
finite derivation tree rooted at $w$. At stage $n$ one lists all words
reachable from $w$ by at most $n$ rewriting steps. If at some stage a
word containing the acceptance symbol $Y$ is found, the enumeration
halts and answers positively.

Thus
\[
        \mathrm{ASSERT}
        =
        \{\langle S,w\rangle \mid S \text{ accepts } w\}
\]
is recursively enumerable.

We now show that the complement of the assertion problem is not
recursively enumerable. The proof is the direct diagonal argument
corresponding to Post's diagonal method for normal systems.

We include the following diagonal proof for two reasons. First, it makes
the paper self-contained at the point where non-recursive enumerability
enters the argument. Second, it exhibits in its simplest form the
Post-style diagonal idea which motivates the present approach: enumerate
all generative systems, apply the system to its own code, and reverse
the resulting assertion.

\begin{theorem}\label{thm:nonassertion-not-re}
The complement of the assertion problem is not recursively enumerable.
That is,
\[
        \overline{\mathrm{ASSERT}}
        =
        \{\langle S,w\rangle \mid S \text{ does not accept } w\}
\]
is not recursively enumerable.
\end{theorem}

\begin{proof}
Assume, for a contradiction, that
$\overline{\mathrm{ASSERT}}$ is recursively enumerable. By the
characterization of recursively enumerable sets by semi-Thue
acceptance, there is a semi-Thue system $T$ with acceptance symbol $Y$
such that, for every semi-Thue system $S$ and every word $w$,
\[
        T \text{ accepts } \langle S,w\rangle
        \quad\Longleftrightarrow\quad
        S \text{ does not accept } w.
\]
In other words, $T$ recognizes precisely the negative instances of the
assertion problem.

Consider now the diagonal set
\[
        K
        =
        \{\langle S\rangle
        \mid
        S \text{ does not accept } \langle S\rangle\}.
\]
The map
\[
        \langle S\rangle
        \longmapsto
        \langle S,\langle S\rangle\rangle
\]
is computable. Since $T$ recognizes
$\overline{\mathrm{ASSERT}}$, it follows that $K$ is recursively
enumerable: on input $\langle S\rangle$, form the code
$\langle S,\langle S\rangle\rangle$ and simulate $T$ on this code.
Thus there exists a semi-Thue system $D$ such that, for every
semi-Thue system $S$,
\[
        D \text{ accepts } \langle S\rangle
        \quad\Longleftrightarrow\quad
        \langle S\rangle\in K.
\]
By the definition of $K$, this means that
\[
        D \text{ accepts } \langle S\rangle
        \quad\Longleftrightarrow\quad
        S \text{ does not accept } \langle S\rangle.
\]

Now take $S=D$. We obtain
\[
        D \text{ accepts } \langle D\rangle
        \quad\Longleftrightarrow\quad
        D \text{ does not accept } \langle D\rangle,
\]
which is impossible. Hence the assumption that
$\overline{\mathrm{ASSERT}}$ is recursively enumerable must be false.
\end{proof}

\subsection{The complement of the word problem}

Recall that the word problem for semi-Thue systems is the set
\[
        \mathrm{WP}
        =
        \{\langle S,u,v\rangle \mid u\to_S^* v\}.
\]
We now show that the complement of the word problem is not recursively
enumerable. The proof uses the result just proved for the assertion
problem.

\begin{theorem}\label{thm:negative-word-problem-not-re}
The complement of the word problem for semi-Thue systems is not
recursively enumerable. That is,
\[
        \overline{\mathrm{WP}}
        =
        \{\langle S,u,v\rangle \mid u\not\to_S^* v\}
\]
is not recursively enumerable.
\end{theorem}

\begin{proof}
We reduce the complement of the assertion problem to the complement of
the word problem.

Let $S=(A,R)$ be a semi-Thue system with acceptance symbol $Y$, and let
$w\in A^*$. We effectively construct a semi-Thue system $S'=(A',R')$
and a word $Z$ such that
\[
        S \text{ accepts } w
        \quad\Longleftrightarrow\quad
        w\to_{S'}^* Z.
\]
Here $Z$ is a new symbol, not in $A$.

Let
\[
        A'=A\cup\{Z\}.
\]
The rules of $S'$ consist of all rules of $S$, together with the
following additional rules:
\[
        Y\to Z,
\]
and, for every $a\in A$,
\[
        aZ\to Z,
        \qquad
        Za\to Z.
\]
We may also include the rule
\[
        ZZ\to Z.
\]
These extra rules have the following purpose. The marker $Z$ can be
created only from an occurrence of the acceptance symbol $Y$. Once $Z$
has appeared, the rules $aZ\to Z$ and $Za\to Z$ erase all neighbouring
letters around it, and $ZZ\to Z$ contracts possible multiple occurrences
of $Z$. Thus any word containing $Y$ can be reduced to the single word
$Z$.

It is now obvious that 
\[
        S \text{ accepts } w
        \quad\Longleftrightarrow\quad
        w\to_{S'}^* Z.
\]
Therefore
\[
        S \text{ does not accept } w
        \quad\Longleftrightarrow\quad
        w\not\to_{S'}^* Z.
\]
The map
\[
        \langle S,w\rangle
        \longmapsto
        \langle S',w,Z\rangle
\]
is effective. Hence, if $\overline{\mathrm{WP}}$ were recursively
enumerable, then $\overline{\mathrm{ASSERT}}$ would also be recursively
enumerable. This contradicts
Theorem~\ref{thm:nonassertion-not-re}. Therefore
$\overline{\mathrm{WP}}$ is not recursively enumerable.
\end{proof}

\section{Non-Termination and Universal Diagonalization}
\label{sec:universal-diagonalization}

\subsection{Why universality matters}
\label{subsec:why-universality-matters}

The proofs in the previous section proceed by assuming the existence of a
semi-Thue system recognizing or deciding the language under consideration.
This is mathematically sufficient for obtaining the desired contradictions.
Conceptually, however, a bare semi-Thue system is merely a rewriting relation:
by itself it has no intrinsic notion of program, input, output, or decision.
These notions become operationally explicit only after introducing a universal
semi-Thue system together with an encoding of systems and inputs and an
interpretation of observable outcomes.

In this sense, the universal semi-Thue system plays the same conceptual role as
the universal Turing machine: it provides an interpreter in which rewriting
becomes computation. Unlike a (deterministic) universal Turing machine, however, the universal
semi-Thue system remains itself an ordinary nondeterministic rewriting system,
and the encoded computation that it simulates is likewise an ordinary
nondeterministic semi-Thue computation. At neither level is the simulated
semi-Thue computation transformed into a deterministic process or endowed
with an external control mechanism.

Thus diagonalization is not performed by first translating semi-Thue systems
into machines with states and then applying the usual machine argument.
Rather, self-reference is realized inside rewriting itself. The universal system
allows a semi-Thue system and an input word to be coded as a single word, and
the diagonal construction then acts on this code. In this way universality is not
merely a technical device for reductions; it is the mechanism which makes
self-reference available in the semi-Thue formalism.

\subsection{A universal semi-Thue interpreter}

Recall the infiniteness problem,
\[
\mathrm{INF}
 =
 \{\langle S,w\rangle \mid
   \text{there exists an infinite derivation from } w
   \text{ in } S\}.
\]
Thus $\langle S,w\rangle\in \mathrm{INF}$ means that the derivation tree
rooted at $w$ has at least one infinite branch. 
Its complementary property is termination:
\[
\mathrm{TERM}
 =
 \{\langle S,w\rangle \mid
   \text{every derivation from } w \text{ in } S \text{ is finite}\}.
\]
That is,
\[
\mathrm{TERM}=\overline{\mathrm{INF}}.
\]

The problem $\mathrm{INF}$ is the semi-Thue analogue of non-halting or
circularity phenomena in machine models. Since semi-Thue derivations are
branching, the existential formulation is essential: non-termination means that
at least one infinite derivation exists.

We now use the fixed universal semi-Thue system $\mathcal{U}$ described in Section~\ref{sec:universal-sT}. Recall
that, for every semi-Thue system $S$ and every word $w$, there is an effective
encoding
\[
        \varphi(S,w)
\]
such that
\[
\text{there exists an infinite derivation from } w \text{ in } S
\]
if and only if
\[
\text{there exists an infinite derivation from } \varphi(S,w) \text{ in } \mathcal{U}.
\]
Equivalently,
\[
        \langle S,w\rangle\in \mathrm{INF}
        \quad\Longleftrightarrow\quad
        \langle \mathcal{U},\varphi(S,w)\rangle\in \mathrm{INF}.
\]

The important point is that both sides of this equivalence are statements about
semi-Thue derivations. The system $\mathcal{U}$ is not an external controller. It is a
semi-Thue system whose derivations simulate the derivations of the encoded
semi-Thue system. Hence the universal construction stays completely inside the
rewriting model.

\subsection{Diagonalization inside the universal system}

We now prove that the infiniteness problem is not recursively enumerable. The
proof is a diagonal argument, but it is important that the construction does
not follow a single branch of a nondeterministic computation. Since acceptance
by a semi-Thue system is existential, an accepting word may occur on one branch
while other branches do not accept. Therefore the constructed system must
enumerate finite levels of derivation trees.

We first record explicitly the finite objects which will be used in the
construction. Let \(T=(A,R)\) be a semi-Thue system and let \(x\in A^*\). Define
the frontiers of the derivation tree of \(T\) from \(x\) by
\[
F_0(T,x)=\{x\}
\]
and
\[
F_{n+1}(T,x)=
\{\,v\mid \exists u\in F_n(T,x)\text{ such that }u\to_T v\,\}.
\]
Thus \(F_n(T,x)\) is the set of all words reachable from \(x\) by exactly
\(n\) rewriting steps. Each \(F_n(T,x)\) is finite, since \(T\) has finitely many
rules and each word has only finitely many one-step successors.

Moreover, these finite frontiers can be effectively written as finite coded
lists. For example, using a separator symbol \(\#\) not in \(A\), a finite frontier
\[
F_n(T,x)=\{w_1,\ldots,w_k\}
\]
may be represented by a word of the form
\[
\# w_1 \# w_2 \# \cdots \# w_k \# .
\]
A rewriting system can then scan such a list, inspect each listed word, and
construct a new list. In particular, it can transform a coded list for
\(F_n(T,x)\) into a coded list for \(F_{n+1}(T,x)\) by applying each rule of
\(T\) at each possible occurrence in each listed word and writing all resulting
words into a new output list.

In the present construction this manipulation of coded frontiers is carried out
uniformly by iteratively writing out the non-deterministic derivation tree using the fixed universal semi-Thue system \(\mathcal{U}\).
The point is that, in
the diagonal construction, the simulated system \(S\) is not built into the
rules of the constructed system. It is supplied as part of the input word.
Thus \(D\) must be able to read a coded rule table and apply its rules to coded
words. The universal system \(\mathcal{U}\) provides exactly this kind of fixed rewriting
interpreter. Given a coded semi-Thue system and a coded word, it can simulate
one rewriting step of the coded system. By using this simulation repeatedly,
and by adding bookkeeping markers, the constructed system can expand an entire
coded frontier into the next coded frontier.

Thus \(D\) will not merely simulate a possible derivation of \(E\) or \(S\).
Rather, the current word of \(D\) will contain coded finite lists representing
frontiers of derivation trees. These lists are part of the actual word being
rewritten, and hence they can be inspected by further rules. In particular,
\(D\) can search a coded frontier for the coded acceptance symbol \(Y\), and it
can test whether a coded frontier is empty.

We now give the diagonal construction.

\begin{theorem}\label{thm:inf-not-re}
The set \(\mathrm{INF}\) is not recursively enumerable.
\end{theorem}

\begin{proof}
Assume, for a contradiction, that \(\mathrm{INF}\) is recursively enumerable.
Then there is a semi-Thue system \(E\), with acceptance symbol \(Y\), such that
for every semi-Thue system \(S\) and every word \(w\),
\[
E \text{ accepts } \langle S,w\rangle
\quad\Longleftrightarrow\quad
\langle S,w\rangle\in \mathrm{INF}.
\]

We construct effectively a semi-Thue system \(D\). On input \(\langle S\rangle\),
the system \(D\) carries out two synchronized frontier enumerations.

First, \(D\) enumerates the frontiers of the derivation tree of \(E\) on the
diagonal input
\[
\langle S,\langle S\rangle\rangle.
\]
That is, it successively writes coded lists representing
\[
F_n\bigl(E,\langle S,\langle S\rangle\rangle\bigr),
\qquad n=0,1,2,\ldots .
\]
After each new frontier has been written, \(D\) scans this finite coded list for
the occurrence of the coded acceptance symbol \(Y\). If such an occurrence is
found, then \(D\) rewrites to an irreducible terminal configuration.
This is done by introducing rules \(aY\rightarrow Y\) and \(Ya\rightarrow Y\) for every \(a\in A\).
Note, that these rules are not part of semi-Thue system \(E\) (or \(S\)).
Recall that an occurrence of \(Y\) does not influence the computation of \(E\) as within \(E\) it has no special meaning such as terminating the computation.

Second, \(D\) enumerates the frontiers of the derivation tree of \(S\) from
\(\langle S\rangle\). Thus it successively writes coded lists representing
\[
F_n(S,\langle S\rangle),
\qquad n=0,1,2,\ldots .
\]
Here the rules of \(S\) are not part of the rules of \(D\); they are read from
the code \(\langle S\rangle\). This is where the use of the universal system \(\mathcal{U}\) is essential. The universal rewriting component reads the coded rule table of
\(S\), applies the encoded rules to each coded word in the current frontier,
and writes the resulting coded words into the next frontier.

After each such expansion, \(D\) checks whether the newly written frontier of
the derivation tree of \(S\) is empty. If the frontier is empty, then all derivations of \(S\)
from \(\langle S\rangle\) have terminated. In that case \(D\) enters a fixed
infinite loop, by introducing a new symbol, say \(\%\), and a rule \(\%\rightarrow \%\%\). 
Thus producing configurations
\[
L_0\to_D L_0\%\to_D L_0\%\%\to_D\cdots
\]
using a simple nonterminating component.

The construction is arranged so that the two tests are performed at each stage
in a fixed order. First \(D\) checks whether the current \(E\)-frontier contains
\(Y\). If it does, \(D\) stops. If not, \(D\) checks whether the current
\(S\)-frontier is empty. If it is empty, \(D\) enters the infinite loop. If
neither event occurs, \(D\) uses the universal component again to construct the
next frontiers and repeats the procedure.

We now verify the behaviour of \(D\). Suppose first that
\[
\langle S,\langle S\rangle\rangle\in \mathrm{INF}.
\]
Since \(E\) recognizes \(\mathrm{INF}\), the system \(E\) accepts
\(\langle S,\langle S\rangle\rangle\). Hence there exists a finite derivation
of \(E\) from \(\langle S,\langle S\rangle\rangle\) to a word containing \(Y\).
Therefore, for some \(n\), the frontier
\[
F_n\bigl(E,\langle S,\langle S\rangle\rangle\bigr)
\]
contains a word in which \(Y\) occurs. Since \(D\) enumerates the frontiers of
the \(E\)-tree level by level and writes them into its current word, \(D\)
eventually finds this occurrence of \(Y\). At that point \(D\) rewrites to an
irreducible terminal configuration.

Moreover, in this case the derivation tree of \(S\) from \(\langle S\rangle\)
has an infinite branch, so its frontiers are never exhausted. Thus the
``empty-frontier'' test for \(S\) never triggers before the accepting symbol
\(Y\) is found during the enumeration of one of the \(E\)-frontiers. Consequently \(D\) has no infinite
derivation from \(\langle S\rangle\) in this case.

Conversely, suppose that
\[
\langle S,\langle S\rangle\rangle\notin \mathrm{INF}.
\]
Then, since \(E\) recognizes \(\mathrm{INF}\), the system \(E\) does not accept
\(\langle S,\langle S\rangle\rangle\). Hence no frontier of the
\(E\)-derivation tree contains a word with \(Y\).

On the other hand, the assumption
\[
\langle S,\langle S\rangle\rangle\notin \mathrm{INF}
\]
means that there is no infinite derivation of \(S\) from \(\langle S\rangle\).
The derivation tree of a semi-Thue system is finitely branching. Hence, by
K\"onig's lemma, if it has no infinite branch, then it is finite. Therefore
there exists \(n\) such that
\[
F_n(S,\langle S\rangle)=\varnothing.
\]
Since \(D\) enumerates and writes the \(S\)-frontiers level by level, it
eventually writes this empty frontier and detects that it is empty. Since the
enumeration of the frontiers by \(E\) has not found \(Y\), the system \(D\) then enters its fixed
infinite loop. Thus \(D\) has an infinite derivation from \(\langle S\rangle\).

We have shown that, for every semi-Thue system \(S\),
\[
\langle D,\langle S\rangle\rangle\in \mathrm{INF}
\quad\Longleftrightarrow\quad
\langle S,\langle S\rangle\rangle\notin \mathrm{INF}.
\]
Now put \(S=D\). We obtain
\[
\langle D,\langle D\rangle\rangle\in \mathrm{INF}
\quad\Longleftrightarrow\quad
\langle D,\langle D\rangle\rangle\notin \mathrm{INF},
\]
which is impossible. Therefore \(\mathrm{INF}\) is not recursively enumerable.
\end{proof}

The essential point of the construction is that the universal system \(\mathcal{U}\) is
used not merely to follow one branch of a simulated nondeterministic system,
but to manipulate coded finite frontiers of whole derivation trees. The finite
frontiers are written into the current word of \(D\), where they can be
examined. Thus an accepting branch of \(E\), if it exists, is eventually found,
and the absence of infinite branches in \(S\) is detected by exhaustion of a
coded frontier. This is why the proof remains entirely inside the semi-Thue
formalism while still handling nondeterminism correctly.

The proof idea of writing out frontiers is somewhat similar to the proof of the undecidability of the emptiness problem for linear-bounded automata (LBA).
Recall, that in the emptiness problem for LBA, we are asked whether a given LBA accepts something.
The undecidability is proven by giving an LBA a finite sequence of configurations of a Turing machine and checking that they correspond to a valid accepting run of the machine.
Our construction has however a crucial difference, the proof relies on self-reference rather than a reduction from another Turing-complete model.

Since $\mathrm{INF}$ is not recursively enumerable, the termination problem
\[
        \mathrm{TERM}=\overline{\mathrm{INF}}
\]
is not decidable. Indeed, if $\mathrm{TERM}$ were decidable, then
$\mathrm{INF}$ would also be decidable, and hence recursively enumerable,
contrary to Theorem~\ref{thm:inf-not-re}.

Thus non-termination is a fundamental source of undecidability for semi-Thue
systems. More importantly for the present paper, the proof shows that this
undecidability arises from universality and self-reference inside
nondeterministic rewriting itself. The universal semi-Thue system does not
replace rewriting by a machine model; it shows that rewriting already contains
the resources needed for diagonalization.

\section{Recognition, Decision, and Observation}

\subsection{Recognition versus decision}

The results of Sections~4 and~5 concern existential properties of derivation
trees. The assertion problem asks whether there exists a finite derivation
reaching a designated symbol, while the infiniteness problem asks whether there
exists an infinite derivation. Both are naturally formulated in terms of the
existence of a branch satisfying a given property.

These notions correspond to recognition rather than decision. Recognition
requires only the existence of a successful derivation witnessing membership.
This is precisely the rewriting-theoretic counterpart of recursive
enumerability.

Decision is a stronger notion. Besides correctness of positive and negative
verdicts, it requires that every maximal derivation eventually reveals one of
them. Thus decision is recognition together with a global requirement that the
observable outcome cannot be avoided by following another branch of the
derivation tree.

The undecidability phenomena established in this paper therefore arise already
at the level of recognition. The additional structure required for decision does
not create undecidability; it merely provides an interpretation of the outcomes
of computation.

\subsection{Verdicts and observation}

The symbols $Y$ and $N$ should not be viewed as computational primitives.
From the viewpoint of a semi-Thue system they are merely ordinary symbols of
the alphabet. Their interpretation as acceptance and rejection is supplied
entirely by an external observer.

The same observation applies to the universal semi-Thue system. The universal
system assigns no distinguished computational meaning to $Y$ or $N$. If the
simulated semi-Thue system derives one of these symbols, then the universal
system merely simulates this derivation. It does not introduce a second level
of acceptance or rejection. Thus universality preserves not only the rewriting
dynamics but also the purely observational nature of verdicts.

This viewpoint is close in spirit to Post's formulation of recursively
enumerable sets by normal systems. A successful computation is not an internal
decision made by the rewriting system itself; it is the external observation
that a word of the required form has appeared in the derivation tree.

\subsection{Computation trees and machine computation}

A Turing machine presents computation as a single sequence of configurations.
Although nondeterministic Turing machines admit branching computations, the
classical presentation of computability theory is fundamentally based on the
deterministic machine, where the computation is represented by one observable
run.

Semi-Thue systems expose a different computational picture. Their natural object
is not a single computation but an entire derivation tree generated by local
rewriting rules. Existential acceptance, reachability, and infiniteness are all
properties of this tree.

From this perspective, the states of a Turing machine may be viewed as a device
for compressing an unbounded computation into a single observable execution.
Semi-Thue systems perform no such compression. They expose the branching
structure directly, making explicit that undecidability is fundamentally a
property of computation trees rather than of machine states.

\subsection{Universality and self-reference}

The universal semi-Thue system plays a role analogous to that of the universal
Turing machine. It provides a fixed interpreter capable of simulating arbitrary
semi-Thue systems from encoded descriptions.

There is, however, an important conceptual distinction. The universal
semi-Thue system is itself an ordinary nondeterministic semi-Thue system, and
the computation that it simulates is likewise an ordinary nondeterministic
semi-Thue computation. At neither level is rewriting replaced by deterministic
control or by an external computational mechanism.

The significance of universality therefore goes beyond providing reductions.
Universality makes self-reference possible entirely within the rewriting model.
A semi-Thue system can simulate the computation of another semi-Thue system,
including one obtained from its own description. The diagonal construction of
Section~5 is carried out entirely inside this framework of nondeterministic
rewriting.

\subsection{Concluding remarks}

The results of this paper suggest that undecidability in semi-Thue systems
should be understood primarily as a consequence of universality and
self-reference rather than as a property of machine states or control
mechanisms. The universal semi-Thue system does not transform rewriting into a
different computational model; it demonstrates that rewriting itself already
contains the resources needed for universal computation and diagonalization.

Decision, by contrast, is an additional observational layer. It arises only
after assigning external meaning to designated verdict symbols and requiring
that every maximal derivation eventually reveals one of them. In this sense,
semi-Thue systems separate two notions that are often intertwined in machine
models: the generation of computations and the observation of their outcomes.

\paragraph{Acknowledgements.} The authors acknowledge the use of ChatGPT, developed by OpenAI, as an editorial and mathematical discussion tool during the preparation of this article. ChatGPT was used to assist with language polishing, structural organization, formulation of explanatory passages, and critical review of draft arguments. In particular, it helped identify points where proofs required further clarification. All mathematical content, final formulations, and responsibility for the results remain entirely with the authors.

\bibliographystyle{plain}
\bibliography{roots.bib}

\end{document}